# A Viscoelastic Deadly Fluid in Carnivorous Pitcher Plants

Laurence Gaume[1]*, Yoel Forterre[2]*

1 CNRS UMR5120-AMAP, Montpellier, France, 2 IUSTI CNRS UMR 6595, Université de Provence, Technopole Château-Gombert, Marseille, France

*Background.* The carnivorous plants of the genus *Nepenthes,* widely distributed in the Asian tropics, rely mostly on nutrients derived from arthropods trapped in their pitcher-shaped leaves and digested by their enzymatic fluid. The genus exhibits a great diversity of prey and pitcher forms and its mechanism of trapping has long intrigued scientists. The slippery inner surfaces of the pitchers, which can be waxy or highly wettable, have so far been considered as the key trapping devices. However, the occurrence of species lacking such epidermal specializations but still effective at trapping insects suggests the possible implication of other mechanisms. *Methodology/Principal Findings.* Using a combination of insect bioassays, high-speed video and rheological measurements, we show that the digestive fluid of *Nepenthes rafflesiana* is highly viscoelastic and that this physical property is crucial for the retention of insects in its traps. Trapping efficiency is shown to remain strong even when the fluid is highly diluted by water, as long as the elastic relaxation time of the fluid is higher than the typical time scale of insect movements. *Conclusions/Significance.* This finding challenges the common classification of *Nepenthes* pitchers as simple passive traps and is of great adaptive significance for these tropical plants, which are often submitted to high rainfalls and variations in fluid concentration. The viscoelastic trap constitutes a cryptic but potentially widespread adaptation of *Nepenthes* species and could be a homologous trait shared through common ancestry with the sundew (*Drosera*) flypaper plants. Such large production of a highly viscoelastic biopolymer fluid in permanent pools is nevertheless unique in the plant kingdom and suggests novel applications for pest control.

## INTRODUCTION

Carnivorous plants live in nutrient-poor soils and have circumvented this shortage of resources by deriving most of their nutrients from the digestion of arthropods captured through a variety of trapping mechanisms [1–4]. The traps are generally formed by highly modified leaves, which can take shapes as diverse [5] as pitfall traps in *Sarracenia*, *Cephalotus* and *Nepenthes*, flypaper-traps in sundews (*Drosera*) and butterworts (*Pinguicula*) or even be very sophisticated devices such as the snap traps of *Dionaea* or the suction bladders of *Utricularia*. All these carnivorous plants secrete a digestive fluid involved in the process of prey digestion [1–4]. Only in the flypaper plants is the fluid also involved in insect capture in addition to its digestive role [1–3]. In these plants, the fluid is secreted by stalked glands in the form of drops of sticky mucilage, where insects are lured and adhere, most of the time irremediably. On the other hand, in pitcher plants such as Nepenthaceae or Sarraceniaceae, the fluid is secreted in far greater quantities in permanent pools within the pitchers (several tens of ml by pitcher compared to the μl-quantities secreted by leaves of flypaper plants); it is never referred to as mucilage and is commonly believed to have as a unique function, prey digestion [1–3].

In *Nepenthes* pitcher plants, prey capture and retention is mainly thought to be fulfilled by the slippery waxy layer which covers the upper inner part of the pitcher in most species [1,6–10], or by the peristome or nectar rim of the pitcher (in *N. bicalcarata* for instance) [11]. However, some *Nepenthes* species lack such specialized surfaces [12] or lose them later in development [13] suggesting that the trapping mechanism of *Nepenthes* pitcher plants is more complex than commonly acknowledged. Moreover, reports of secretion of wetting agents [3] or viscous substances [14] in some species point to other potential roles of the digestive fluid.

Here we focus on *N. rafflesiana*, one of the most widespread species of the genus in northern Borneo [12,15] (Fig. 1a). It is common in heath forests and has one of the richest prey spectra of any species in the genus [13,16]. However, in this species the waxy layer is a variable character and is probably of weak adaptive significance since comparison of waxy traps and non-waxy traps did not show any difference in their amount of prey captured [13]. In contrast, the plant secretes a large amount of slimy fluid, which forms sticky filaments when rubbed between the fingers (pers. observ.). Moreover, field observations on insects fallen in the pitchers reveal that they sink and are easily drawn within the pitchers [13]. This could suggest that the physical properties of the fluid are implicated in insect trapping in this species. A slightly lower surface tension (compared to water) has been observed in the fluid of *Sarracenia* pitcher plants [17] and was suspected to be part of the trapping in *Nepenthes* by Juniper and co-authors [3] but to our knowledge, no measure of fluid surface tension has been conducted on any species of *Nepenthes* pitcher plants. Moreover, up to now, the rheological properties of the fluid, which govern how a fluid moves under forces, and their possible role in insect capture have never been investigated. We thus focused our study on the digestive fluid of *N. rafflesiana* and first tested whether the fluid alone was able to retain insects by comparing retention of insects thrown into glass vials filled with water or pure digestive fluid. Then, to determinate which physical properties of the digestive

. . . . . . . . . . . . . . . . . . . . . . . . . . . . . . . . . . . . . . . . . . . . . . . . . . . . . . . .

Funding: The work was supported by a grant "Young researcher and innovative project" from the University Montpellier II and by the French national agency of research (ANR) "Young researcher program: Interaction fluide-structure chez les plantes". The study funders played no role in designing the study, data collection/analysis or preparation of the manuscript.

Competing Interests: The authors have declared that no competing interests exist.

* To whom correspondence should be addressed. E-mail: lgaume@cirad.fr (LG); Yoel.Forterre@polytech.univ-mrs.fr (YF)

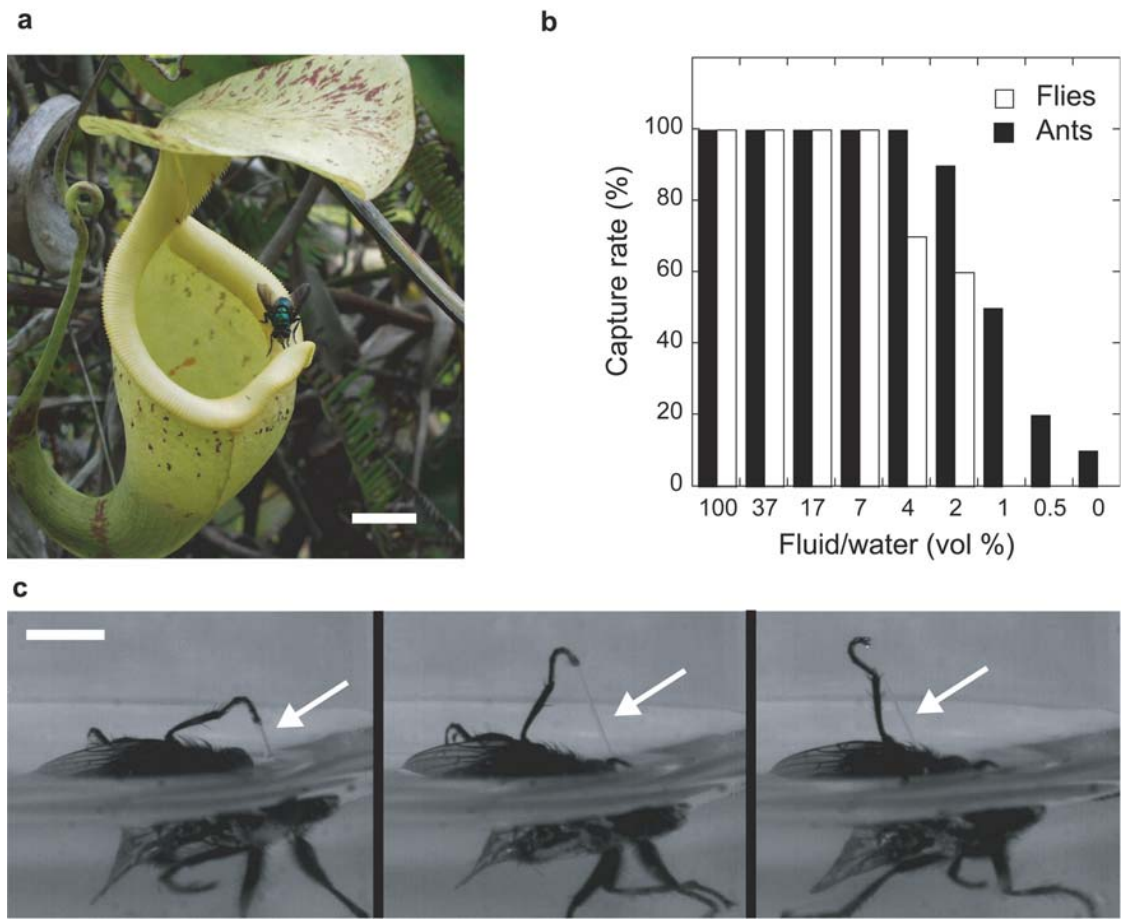

**Figure 1. Capture property of the digestive fluid of *Nepenthes rafflesiana*.** (**a**) Pitcher of *N. rafflesiana* showing a *Calliphora* fly collecting extrafloral nectar in a perilous position (Brunei). (**b**) Capture rate of *C. vomitoria* flies (white) and *L. humile* ants (black) thrown into glass vials filled with pure digestive fluid (100%), water (0%) and intermediate dilutions (digestive fluid from a mixing of 7 pitchers of 7 individual plants). (**c**) Dynamical sequence of a fly (*C. vomitoria*) after falling into the digestive fluid, showing a viscoelastic liquid filament attached to its leg (arrows). Time between frames: 80 ms. Scale bars, 1 cm (**a**); 3 mm (**c**).
doi:10.1371/journal.pone.0001185.g001

fluid was implicated in retention, we compared surface properties (surface tension, static wetting) and rheology (viscosity, elasticity) of pure fluid, water and intermediate dilutions. Our study unveils the peculiar viscoelastic properties of the digestive fluid of *N. rafflesiana* and its crucial role in prey capture.

## RESULTS

### Role of the digestive fluid in insect trapping

The fluid was collected from young and newly opened pitchers of *N. rafflesiana*. The insects (ant workers and flies) were chosen because they represent non-flying and flying insects and are part of the two insect orders (Hymenoptera and Diptera, respectively) most frequently captured by the pitcher plants [3,16]. We first confirmed that flies (*Drosophila melanogaster*, *Calliphora vomitoria*) and ants (*Linepithema humile*) escaped easily from water (successes: 10/10, 10/10 and 9/10 respectively, see Fig 1b for the two last insects). In water, flies typically succeeded in taking off and flying away in a few seconds (high-speed movies: Video S1, S2), whereas ants succeeded in swimming and climbing up the glass walls of the vials. These observations are in strong contrast with observations of the same insects thrown into the digestive fluid of *N. rafflesiana*. In this fluid, insects were never able to escape during the 5 minutes observations of the tests (successes: 0/10 for the three insect types, Fig 1b, Video S3, S4). High-speed videos revealed that flies were wetted by the digestive fluid and were unable to move their wings and extract their legs, which were retained by sticky filaments typical of complex fluids such as mucus or saliva (Fig. 1c, Video S4). Surprisingly, we observed that the trapping efficiency of the digestive fluid remained maximal even when the fluid was highly diluted by water. The capture rate started to drop only for fluids diluted by more than 95% (Fig. 1b). These results show that the digestive fluid of *N. rafflesiana* has on its own, and outside of any effect of the pitcher wall, a very high retention capability. It is important to note that insects, when experimentally removed from the fluid 5 minutes after being trapped, soon recovered their normal activity. This suggests that the capture property of the digestive fluid does not result from a rapid chemical attack but is primarily physical in nature.

### Surface tension and wetting properties are not involved in insect trapping

Since insects greatly rely on their hydrophobic body surfaces and on the surface tension of water to sustain their weight and move at liquid interfaces [18], one can wonder whether surface physical properties of the digestive fluid (low surface tension, wetting properties) could explain insect retention. Such surface mechanisms could also explain why highly diluted fluids are still efficient in catching insects in *Nepenthes*. To check this, we first measured precisely the surface tension $\sigma$ of the pure digestive fluid and found it to be very similar to that of de-ionized water measured in the same condition ($\sigma_{fluid} = 0.0725 \pm 0.0024$ N.m$^{-1}$, n = 113, 12 samples from 12 young pitchers; $\sigma_{water} = 0.0720 \pm 0.0012$ N.m$^{-1}$, n = 13; t-test for unequal variances, $t = -1.15$, $p = 0.26$). Moreover, while observations on moving insects showed that the pure digestive fluid wet flies and ants once they began to move in the fluid, observations on paralyzed insects showed that the pure digestive fluid did not spontaneously wet insects. This suggests that wetting of insects occurs dynamically rather than statically, probably because of fluid viscosity. More quantitative data were obtained by measuring the quasi-static advancing contact angles of the fluid (same mixing as in the retention experiment) over model surfaces using the sessile drop method [19]. Contact angle of the digestive fluid on a polystyrene surface (plastic Petri dish) was high (>90°) meaning that the fluid does not wet the surface, and did not vary when diluted by water (n = 80). Similar results (n = 30) were obtained using Teflon surfaces and surfaces coated with Lycopodium powder, which mimics the super-hydrophobic nature of insects' cuticle [20,21]. Therefore, surface effects are unlikely to explain the capture properties of the digestive fluid.

## The viscoelastic properties of the fluid as the main trapping device

Besides forces generated by surface tension, insects struggling on the liquid surface have to overcome hydrodynamic drag forces to escape from the fluid [22]. To estimate the resistance of the fluid to insect movements, we first measured the shear viscosity $\eta$ of the digestive fluid (the coefficient of proportionality between the shear stress and the shear rate in a simple shear flow [23,24]) using a Brookfield DVII low-viscosity cylindrical Couette rheometer. We found the digestive fluid to be shear-thinning, its viscosity decreasing with the applied shear rate $\dot{\gamma}$ (Fig. 2a, same fluid mixing as in the retention experiment). At a shear rate corresponding to the flies' typical stroke in the fluids ($\dot{\gamma} \sim V/d \sim 40$ s$^{-1}$, where $V \sim 20$ cm.s$^{-1}$ is the typical leg velocity obtained from the high-speed videos and $d \sim 0.5$ cm is the typical leg size), the shear viscosity of the pure fluid was rather low, about 15 times the viscosity of water ($\eta = 15.01 \pm 4.36$ mPa.s, n = 21). However, when insects move, the digestive fluid is not only sheared but also stretched. For simple liquids such as water, resistance to both shearing and stretching are equivalent and given by the same value of the viscosity (except for a factor 3 due to geometrical effects [24]). However, for the digestive fluid, the occurrence of long-lived filaments (Fig. 1c) suggests that the resistance to extensional flows, the so-called extensional viscosity [24], is actually much larger than the classical shear viscosity. Such an effect is typical of complex fluids composed of long-chain polymers, and results from the additional elastic stresses needed to stretch the molecules [23–25]. To quantify this elastic behavior, we estimated the elastic relaxation time $\lambda$ (the time required for the elastic structures in the fluid to relax) and apparent extensional viscosity $\eta_E$ (the ratio between the normal stress and the extensional rate in a uniaxial extensional flow) of the digestive fluid in a controlled capillary break-up extensional geometry [26] (Fig. 2b, Methods). We found that the apparent extensional viscosity strongly increased when the fluid was stretched, being $10^4$ times larger than the shear viscosity at typical shear rates and strains imposed by insect motions (Fig. 2). Systematic experiments performed on the pure fluid, water and intermediate dilutions demonstrated that elasticity is the only property that can explain continued trapping efficiency at low fluid concentration. While the shear viscosity became similar to water at concentrations for which the capture rate was still maximal, the elastic relaxation time and the extensional viscosity remained high even for highly diluted fluids (Fig. 3).

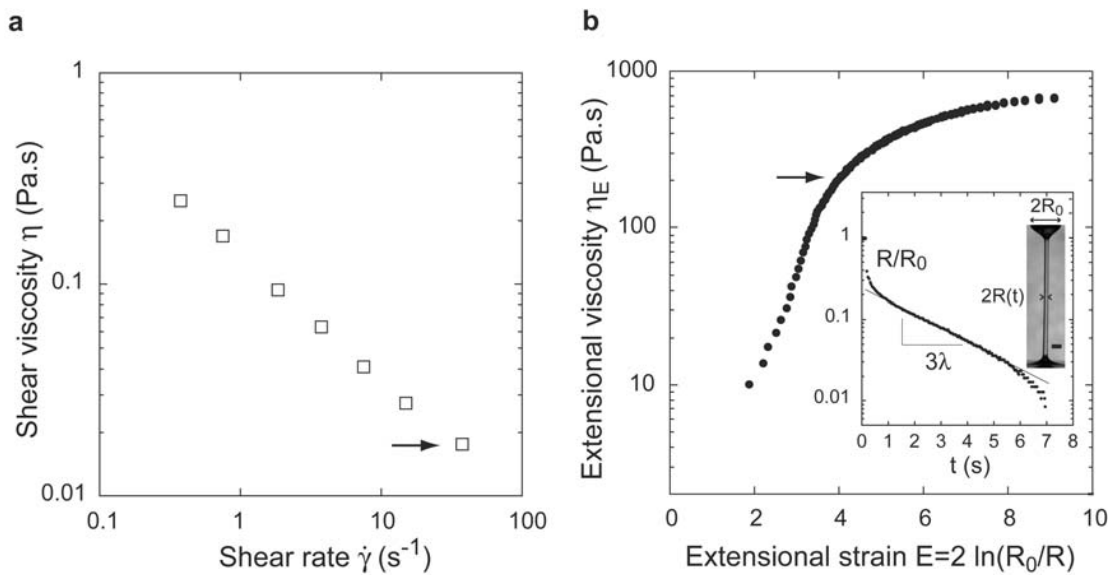

Figure 2. Viscosity and elasticity of the pure digestive fluid of *N. rafflesiana*. (a) Shear viscosity $\eta$ as a function of the shear rate $\dot{\gamma}$ (same mixing of fluids as in retention experiments). (b) Typical transient extensional viscosity $\eta_E$ as a function of the extensional strain $E$, obtained from the thinning dynamics of a liquid filament in a capillary break-up geometry (see inset, the solid line materializes the elasto-capillary exponential regime giving the elastic relaxation time $\lambda$). The high increase of the extensional viscosity with strain is a signature of fluid elasticity. Arrows in (a) and (b) indicate the typical values of shear viscosity and extensional viscosity corresponding to insect motion in the fluid ($\dot{\gamma} \sim 40$ s$^{-1}$ and $E \sim \dot{\gamma}\tau \sim 4$ with $\tau \sim 0.1$ s the typical time scale for insect motion). Scale bars, 1 mm.
doi:10.1371/journal.pone.0001185.g002

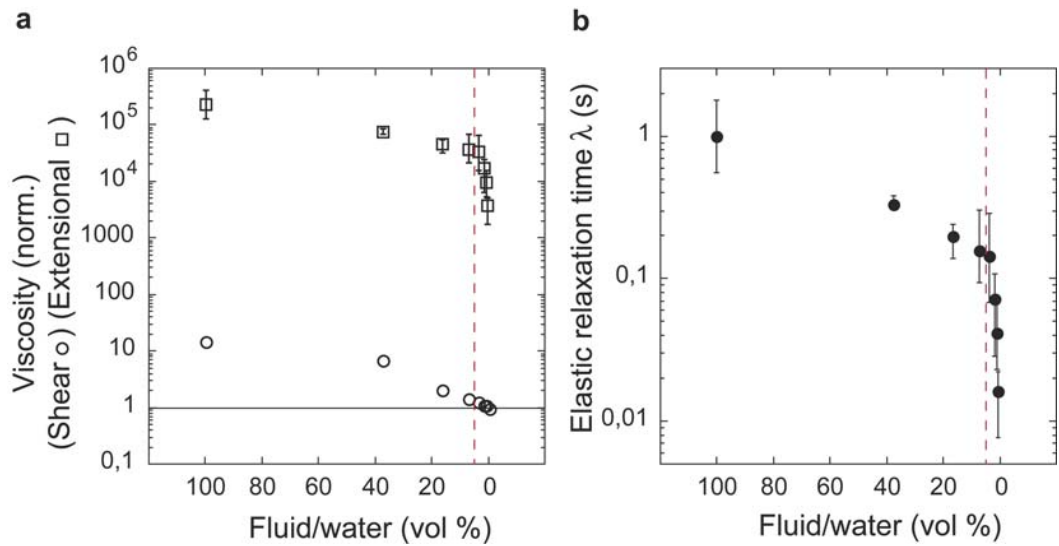

Figure 3. Effect of dilution on the viscoelastic properties of the digestive fluid. The red dotted vertical line materializes the abrupt transition in capture rate for insects (see Fig. 1b). (a) Extensional viscosity (white squares) and shear viscosity (white circles) of the pure fluid and diluted solutions, normalized by the shear viscosity of water (solid line, $\eta^{water} = 0.0012$ Pa.s, measured in the same condition). Shear viscosity is plotted for $\dot{\gamma} = 40$ s$^{-1}$ and extensional viscosity is plotted for $E = 4$, the typical shear rate and extensional strain of the fluid imposed by insect motion respectively. (b) Elastic relaxation time $\lambda$ of the pure fluid and diluted solutions (M±min-max of 10 fluids from 10 pitchers of 7 plants).
doi:10.1371/journal.pone.0001185.g003

## Trapping efficiency is conditioned by both fluid viscoelasticity and insect dynamics

The previous result shows a correlative relationship between the capture rate of insects and the viscoelastic properties of the fluid estimated by its extensional viscosity. However it does not provide any mechanistic explanation of insect trapping. One important parameter, which characterizes the dynamics of viscoelastic fluids, is the Deborah number [23,24]. The Deborah number is the ratio of the fluid elastic relaxation time to the typical time scale of fluid movement. For small Deborah numbers, the time scale of fluid movement is large compared to the relaxation time of elastic forces: the fluid thus behaves like a simple viscous fluid. For large Deborah numbers, the fluid movement is too fast for elastic forces to relax: in this case the fluid behaves like an elastic material. When insects struggle in the pitcher fluid, insect movements control the time scale of the flow. In order to test whether the capture rate could be linked to the Deborah number, we calculated for each fluid dilution the Deborah number ($De$) as the ratio of the fluid elastic relaxation time $\lambda$ (each $\lambda$ was obtained from the mean values for each fluid dilution in Fig. 3b) to the typical half-period of the swimming stroke of insects measured with a high-speed camera $\tau$ ($\tau_{flies}$ = 0.09±0.02 s [from 8 flies over n = 40 periods], $\tau_{ants}$ = 0.12±0.02 s [from 10 ants over n = 50 periods], no systematic dependence of the swimming stroke of insects according to fluid dilution was detected). As shown by the results of a logistic regression (Fig. 4), the capture rate increased significantly with the Deborah number [$\chi^2$ = 146.40, $p<0.0001$] and was significantly higher for ants than flies [effect of the insect type: $\chi^2$ = 24.29, $p<0.0001$], while the two fitted lines did not differ significantly for their slopes (interaction Deborah number * Insect type not significant [$\chi^2$ = 3.45, $p$ = 0.063]). The important result is that the abrupt transition in capture rate occurred when the Deborah number became inferior to 1, i.e. when the elastic relaxation time became inferior to the typical half-period of the swimming stroke of insects (Fig. 4). This suggests that trapping occurs when the elastic forces created by insect movements have no time to relax.

## DISCUSSION

From cellular cytoplasm to animal mucus and plant mucilage, viscoelastic mucilaginous fluids are often involved in important functions ranging from cell mechanical properties [27], transport in lung flows [28], attachment and locomotion [29] or defense [30] in limbless animals, to water storage, food reserves, seed germination [31] and nutrient uptake by roots [32] in plants. Here we discovered that the fluid of the pitcher plant *Nepenthes rafflesiana* is viscoelastic and plays a crucial role in the trapping function of the carnivorous plant. Our study challenges the dogma according to which the pitchers of *Nepenthes* are passive pitfall traps relying on surface structure and gravity to catch insects [3,6]. The elastic behavior of the fluid in species such as *N. rafflesiana*, causing unsuspected highly retentive forces stimulated by insect movement, would warrant the inclusion of *Nepenthes* among other active carnivorous species such as the Venus flytrap whose conspicuous trapping mechanism is also based on elastic forces [33].

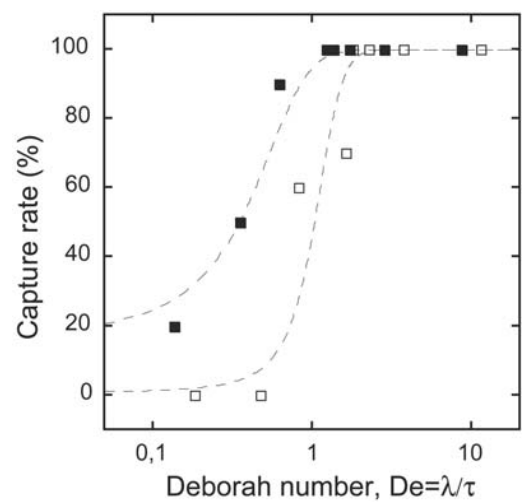

Figure 4. Capture rate of insects as a function of the Deborah number. Flies (empty squares), ants (filled squares). The Deborah number ($De$) is defined as the ratio of the fluid elastic relaxation time $\lambda$ (see Fig. 3b) to the typical half-period of the swimming stroke of insects in the fluid $\tau$. For each insect category, the capture rate decreases rapidly when $De<1$, suggesting that trapping occurs when the elastic forces created by the insect's movements have no time to relax.
doi:10.1371/journal.pone.0001185.g004

Outside of any effect of the pitcher walls, the pitcher fluid alone is able to retain insects (flies and ants) in the trap of *N. rafflesiana* and this striking effect is due to physical properties. The magnitude of the fluid effect is not an artifact due to laboratory conditions or to the choice of insect species. We obtained results of the same order of magnitude in field experiments conducted in the same conditions of pitcher age and insect prey (same families and sizes, unpublished data). The elastic forces are likely to be the key force of retention of insects in the fluid of *N. rafflesiana*. First, the fluid did not spontaneously wet insects and its surface tension was shown to be similar to that of water from which insects can escape easily. Therefore, surface properties of the fluid are unlikely to be involved in trapping in *N. rafflesiana*. Moreover, the shear viscosity of the fluid at typical shear rates imposed by insect movements was rather low, and became similar to that of water at dilutions for which the capture rate was still maximal. We therefore suggest that the shear viscosity of the digestive fluid is unlikely to explain the retention of insects, although further investigation using standardized simple (non elastic) fluids would be helpful to precisely quantify a possible role of shear viscosity in trapping. By contrast, the digestive fluid exhibits a strong extensional viscosity, which appears to be several orders of magnitude larger than the classical (shear) viscosity. This is a clear signature of elasticity in complex fluids. Unlike the shear viscosity, the extensional viscosity of the digestive fluid remains high even when the fluid was highly diluted by water. Such a large extensional viscosity means that the digestive fluid offers a large resistance to stretching or squeezing flows, hence penalizing insects when they try to escape from the fluid or to climb on the pitcher wall. The strong correlation we found between the capture rate of insects and the Deborah number supports the hypothesis that trapping occurs when the elastic forces created by insect movements have no time to relax. It is also compatible with recent models of locomotion in viscoelastic fluids showing that propulsion is much less efficient at high Deborah numbers [34]. This fluid is therefore lethal to insects, which most of the time, once fallen in the pitcher, panic and exhibit quick movements. Their sole chance of escape would be to move slowly. This is perhaps the strategy adopted by the nepenthebiont crab spider *Misumenops nepenthicola*, which lives and reproduces in the trap of *N. rafflesiana* [14,15] and can enter and escape safely from the fluid from which it removes prey items for its own consumption.

Our result undermines the claim that the pitcher surfaces are the main component of the trapping mechanism in *Nepenthes* [3,6–10]. In *N. rafflesiana*, the slippery waxy surface of the pitchers was shown to play a minor role in the trapping function of the plant, being even a variable character of weak adaptive significance [13]. In contrast, the viscoelastic and retentive properties of the fluid are probably crucial for this tropical plant often submitted to high rainfall regimes and great variation in fluid concentration, since they persist at high dilutions by water, thereby allowing insect trapping during rainy seasons. Our results show that even a fluid with a shear viscosity almost similar to that of water might be elastic enough to capture insects. We therefore suggest that such a cryptic property, demonstrated here for *Nepenthes rafflesiana*, may also apply to other species and be more frequent than commonly acknowledged within the *Nepenthes* genus. Already, casual observations of insects attached to the inner pitcher wall, as if they were glued by plant secretions, were made in *N. inermis* [14]. In that species, a highly viscid fluid was suspected to retain dead prey in event of flooding during rain [35]. It is important to note that the pitchers of *N. inermis* lack a waxy zone. It is also the case of *N. eymae*, *N. aristolochioides*, *N. talagensis*, *N. dubia* and *N. jacquelinae*, whose trapping fluid has been reported to be viscous in touching [14]. *Nepenthes inermis* was reported to be (under the name of *N. bongso*) specialised in trapping midges [36]. Similar patterns were observed for *N. aristolochioides*, while *N. jacquelinae* was observed to trap essentially flying preys of bigger sizes [14]. The diversity of characters involved in trapping such as the viscoelastic fluid and the slippery surfaces in Nepenthes pitcher plants would certainly warrant comparative analyses in an evolutionary context. Moreover, ecological studies would help to clarify the selective pressures that have led to the evolution of different trapping strategies in this carnivorous genus.

The exact composition of the viscoelastic fluid remains to be studied. However, we can suppose that the fluid is composed of polysaccharides, as these macromolecules are the main component of mucilages in plants [31]. The fluid of *Nepenthes* could exhibit a composition close to the acidic polysaccharide mucilage [37] of the related sundew or flypaper plant [5]. Thus the structure and associated viscoelastic properties of this mucilage could constitute homologous traits shared through common ancestry with these flypaper plants. Its abundant production in external pools is however to our knowledge unique in plants. Therefore, this fluid could serve as a model for applications in pest control, such as the confection of insect glues or insecticide sprays that avoid the problem of drop bouncing on plants [38]; and, as several other plant mucilages, it could be used in the pharmaceutical and food industries for different purposes [31].

## MATERIALS AND METHODS

### Plant specimens
All measurements were performed in the laboratory on pitchers of plants grown in greenhouses in Montpellier (CEFE-CNRS) at 25–30°C and 90% humidity. Complementary measurements of surface tension and viscosity were carried out in the fields (heath forest of Brunei Darussalam, Borneo) and gave similar results.

### Retention experiments
For the retention experiment, about 100 *Calliphora vomitoria* larvae were bred at 27°C until adult emergence. The flies *Drosophila melanogaster* were bred on a nutritive substrate and the ants *Linepithema humile* were collected on the grounds of the campus of the University in Marseille (Polytech'Marseille DME, Technopole Château-Gombert). Pure digestive fluid of seven newly opened pitchers from seven plants was collected and mixed in a glass vial in the laboratory at 25°C. Nine vials (each filled with 50 ml) were prepared from this mixture using different solutions, respectively 100% pure fluid, 37%, 17%, 7%, 4%, 2%, 1%, 0.5% fluid and distilled water. The experiment consisted in drawing a fly into a soft tube, blowing it onto the solution in a given vial and observing it during five minutes. Ten trials, each using a different fly, were carried out for each of the nine fluid solutions. The capture rate was defined as the percentage of flies that were still retained within the fluid solution after 5 minutes. All the flies that did not escape within five minutes eventually died within the fluid. Similar experiments were conducted with the ants.

### Surface tension measurements
Interfacial surface tension between air and pure digestive fluids (12 fluids from 12 opening pitchers of 6 different plants) was measured using the pendant drop method [39] at ambient temperature (25°C). Millimeter-sized drops were produced from cleaned Pasteur pipettes and photographed using a high-resolution (3008*2000 pixels) digital camera (Nikon D70, AF Micro Nikkor 105 mm lens). Images (10 per pitcher) were post-processed using Image J (http://rsb.info.nih.gov/ij/) and Matlab® softwares in order to compute each drop's equatorial diameter $D$ and diameter $d$ at the distance $D$ from the bottom of the drop. The interfacial surface tension was then computed from $\sigma = \rho g D^2/H$, where $\rho$ is the density of the digestive fluids ($\rho = 1004$ kg.m$^{-3}$, averaged over the 12 pitchers), $g = 9.81$ m.s$^{-2}$ is the gravity and $H$ is a shape parameter depending on the ratio $d/D$ [39]. Since the digestive fluid is viscoelastic, special care was taken to ensure that the drop's interface was in equilibrium before making measurements. Our method of measurement was reliable since the water/air surface tension ($\sigma = 0.0720$ N.m$^{-1}$) we obtained was equal to the reference value given at 25°C ($\sigma = 0.07197$ N.m$^{-1}$) [40].

### Extensional rheometry

Fluid elasticity of pure and diluted fluids (10 fluids from 10 young pitchers from the seven plants used in the previous analyses) was investigated using capillary break-up extensional rheometry [26]. To do so, an axial step strain (step time scale $\delta t = 0.056$ s, final separation 12.5 mm) was applied to the fluid by rapidly vertically removing a thin rod (radius $R_0 = 1.5$ mm) from a small sample of fluid, thus creating an elongated liquid filament. The subsequent capillary thinning and break-up dynamics of the filament were recorded at a high spatial and temporal resolution (6.25 μm/pixel, up to 500 frames/s) using a high-speed camera (IDT Monochrome) mounted on a stereomicroscope (Leica MZ16). Videos (n = 5 for each tested fluid) were then post-processed using Image J and Matlab® software in order to compute the time evolution of the midpoint filament's radius $R(t)$. In all experiments, gravity was small compared to capillary forces (Bond number $Bo = \rho g R_0^2/\sigma \sim 0.3$) and the filament's relaxation time scale was large compared to the inertial time scale $t_i = \sqrt{\rho R_0^3/\sigma}$ and viscous time scale $t_v = \eta R_0/\sigma$. In this case, the dynamics of the filament is mainly controlled by the equilibrium between capillarity, which drives thinning, and elasticity, which opposes thinning [25]. For model elastic fluids, the midpoint radius is then predicted to relax exponentially with time $R(t) \sim \exp(-t/3\lambda)$, where $\lambda$ is the longest relaxation time scale of the internal elastic forces [24–25]. From the $R(t)$ curve, we computed the elastic relaxation time $\lambda = (1/3)\int_0^\infty tR(t)dt / \int_0^\infty R(t)dt$, the transient (apparent) extensional viscosity $\eta_E = -3\sigma/(14.1 dR/dt)$ and the total extensional (Hencky) strain $E(t) = \int_0^t (dE/dt')dt' = \int_0^t -(2/R)(dR/dt')dt' = 2\ln(R_0/R)$ [26].

### SUPPORTING INFORMATION

**Video S1** This high-speed video (500 frames/s, total time = 0.4 s) shows the fall and escape of a fly (*Calliphora vomitoria*) thrown into water on its ventral surface (QuickTime, 2.8 MB).
Found at: doi:10.1371/journal.pone.0001185.s001 (2.89 MB MOV)

**Video S2** This high-speed video (500 frames/s, total time = 1.6 s) shows the fall and escape of a fly (*Calliphora vomitoria*) thrown into water on its dorsal surface (QuickTime, 3.4 MB).
Found at: doi:10.1371/journal.pone.0001185.s002 (3.53 MB MOV)

**Video S3** This high-speed video (500 frames/s, total time = 1.6 s) shows the fall and retention of a fly (*Calliphora vomitoria*) thrown into the digestive fluid on its ventral surface. The motion of the fly involves both shearing and stretching of the fluid at large Deborah numbers (10<De<100), thereby inducing high retentive elastic forces (the Deborah number De is defined the ratio of the elastic relaxation time of the fluid to the typical half period of the swimming stroke of insects, see text) (QuickTime, 3.7 MB).
Found at: doi:10.1371/journal.pone.0001185.s003 (3.92 MB MOV)

**Video S4** This high-speed video (500 frames/s, total time = 1.5 s) shows the fall and retention of a fly (*Calliphora vomitoria*) thrown into the digestive fluid on its dorsal surface. The fly is wetted by the digestive fluid and is unable to move its wings and extract its legs, which are retained by sticky filaments typical of complex fluids such as mucus or saliva (QuickTime, 8.7 MB).
Found at: doi:10.1371/journal.pone.0001185.s004 (9.16 MB MOV)

### ACKNOWLEDGMENTS

J.-J. Labat (French Conservatory of Carnivorous Plants, Peyrusse-Massas) is greatly acknowledged for having supplied the plants used in this study. The authors would like to thank J. Dumais, E. Jousselin, B. Meyer-Berthaud and D. McKey for their critical reading and helpful comments on the manuscript.

### Author Contributions

Conceived and designed the experiments: LG YF. Performed the experiments: LG YF. Analyzed the data: LG YF. Wrote the paper: LG YF.

### REFERENCES

1. Darwin C (1875) Insectivorous plants. London: John Murray. 376 p.
2. Lloyd FE (1942) The carnivorous plants. Waltham, Mass., US: Chronica Botanica Co. 352 p.
3. Juniper BE, Robins RJ, Joel DM (1989) The Carnivorous Plants. London: Academic Press. 353 p.
4. Ellison AM, Gotelli NJ, Brewer JS, Cochran-Stafira DL, Kneitel JM, et al. (2003) The evolutionary ecology of carnivorous plants. Advances in Ecological Research 33: 1–74.
5. Albert VA, Williams SE, Chase MW (1992) Carnivorous plants: phylogeny and structural evolution. Science 257: 1491–1495.
6. Juniper BE, Burras J (1962) How pitcher plants trap insects. New Scientist 13: 75–77.
7. Gaume L, Gorb S, Rowe N (2002) Function of epidermal surfaces in the trapping efficiency of *Nepenthes alata* pitchers. New Phytologist 156: 479–489.
8. Gaume L, Perret P, Gorb E, Gorb S, Labat J-J, Rowe N (2004) How do plant waxes cause flies to slide? Experimental tests of wax-based trapping mechanisms in three pitfall carnivorous plants. Arthropod Structure and Development 33: 103–111.
9. Gorb E, Haas K, Henrich A, Enders S, Barbakadze N, et al. (2005) Composite structure of the crystalline epicuticular wax layer of the slippery zone in the pitchers of the carnivorous plant *Nepenthes alata* and its effect on insect attachment. J Exp Biol 208: 4651–4662.
10. Riedel M, Eichner A, Meimberg H, Jetter R (2007) Chemical composition of epicuticular wax crystals on the slippery zone of pitchers of five *Nepenthes* species and hybrids. Planta 225: 1517–1534.
11. Bohn HF, Federle W (2004) Insect aquaplaning: *Nepenthes* pitcher plants capture prey with the peristome, a fully wettable water-lubricated anisotropic surface. Proc Natl Acad Sci U S A 101: 14138–14143.
12. Cheek M, Jebb M (2001) Nepenthaceae. In: Flora Malesiana-Serie I-Seed Plants, Leiden, the Netherlands: Publication Department of the National Herbarium Nederland 15: 1–164.
13. Di Giusto B, Guéroult M, Rowe N, Gaume L (in press) The waxy surface in *Nepenthes* pitcher plants: variability, adaptive significance and developmental evolution. In: Gorb S, ed. Functional Surfaces in Biology. Berlin: Springer.
14. Clarke C (2001) *Nepenthes* of Sumatra and Peninsular Malaysia. Kota Kinabalu, Sabah, Malaysia: Natural History Publications.

15. Clarke C (1997) *Nepenthes* of Borneo. Kota Kinabalu, Sabah, Malaysia: Natural History Publications.
16. Moran JA (1996) Pitcher dimorphism, prey composition and the mechanism of prey attraction in the pitcher plant *Nepenthes rafflesiana* in Borneo. Journal of Ecology 84: 515–525.
17. Hepburn JS, Jones FM, St John EQ (1927) Observations on the pitcher liquor of the Sarraceniaceae. Wagner Free Inst Sci 11: 35–49.
18. Bush JWM, Hu DL (2006) Walking on water: biolocomotion at the interface. Ann Rev Fluid Mech 38: 339–369.
19. Gu Y (2006) Contact angle measurement techniques for determination of wettability. In: Somasundaran P, ed. Encyclopedia of Surface and Colloid Science. New York: Marcel Dekker Inc. 1525 p.
20. Holdgate MW (1955) The wetting of insect cuticles by water. J Exp Biol 32: 591–617.
21. Gorb SN, Kesel A, Berger J (2000) Microsculpture of the wing surface in *Odonata*: evidence for cuticular wax covering. Arthropod Structure and Development 29: 129–135.
22. Vogel S (1994) Life in Moving Fluids-The physical biology of flows (second edition), Princeton: Princeton University Press. 467 p.
23. Larson RG (1999) The structure and rheology of complex fluids. New York: Oxford University Press. 663 p.
24. Bird RB, Amstrong RC, Hassager O (1987) Dynamics of Polymeric liquids, vols 1 New York: Wiley. 672 p.
25. Entov VM, Hinch EJ (1997) Effect of a spectrum of relaxation times on the capillary thinning of a filament of elastic liquid. J Non-Newtonian Fluid Mech 72: 31–54.
26. Rodd LE, Scott TP, Cooper-White JJ, McKinley GH (2005) Capillary break-up rheometry of low-viscosity elastic fluids. Appl Rheo 15: 12–27.
27. Balland M, Desprat N, Asnacios A, Richert A, Gallet F (2006) Universal power laws in creep function and viscoelastic complex modulus microrheology on living cells. Phys Rev E74: 021911.
28. Grotberg JB (1994) Respiratory fluid mechanics and transport processes. Ann Rev Biomed Eng 3: 421–457.
29. Denny MW (1980) The role of gastropod pedal mucus in locomotion. Nature 285: 160–161.
30. Gaume L, Shenoy M, Zacharias M, Borges R (2006) Co-existence of ants and an arboreal earthworm in a myrmecophyte of the Indian Western Ghats: anti-predation effect of the earthworm mucus. J Trop Ecol 22: 341–344.
31. Morton JF (1990) Mucilaginous plants and their uses in medicine. Journal of Ethnopharmacology 29: 245–266.
32. Read DB, Bengough AG, Gregory PJ, Crawford JW, Robinson D, et al. (2003) Plant roots release phospholipid surfactants that modify the physical and chemical properties of soil. New phytologist 157: 315–326.
33. Forterre Y, Skotheim JM, Dumais J, Mahadevan L (2005) How the Venus flytrap snaps. Nature 433: 421–425.
34. Lauga E (2007) Propulsion in a viscoelastic fluid. Phys Fluid 19: 083104.
35. Salmon B (1993) Some observations on the trapping mechanisms of *Nepenthes inermis* and *N. rhombicaulis*. Carnivorous Plant Newsletter 23: 101–114.
36. Kato M, Hotta M, Tamin R, Itino T (1993) Inter- and intra-specific variation in prey assemblages and inhabitant communities in *Nepenthes* pitchers in Sumatra. Tropical Zoology 6: 11–25.
37. Gowda DC, Reuter G, Schauer R (1982) Structural features of an acidic polysaccharide from the mucin of *Drosera binata*. Phytochemistry 21: 2297–2300.
38. Bergeron V, Bonn D, Martin JY, Vovelle L (2000) Controlling droplet deposition with polymer additives. Nature 405: 772–775.
39. Drelich J, Fang C, White CL (2006) Interfacial tension measurement in fluid-fluid systems. In: Somasundaran P, ed. Encyclopedia of Surface and Colloid Science. New York: Marcel Dekker Inc. 2966 p.
40. Lide DR (2004) Handbook of Chemistry and Physics $85^{st}$ ed. Boca Raton: CRC Press. 2656 p.